  \documentstyle[12pt]{article}
\catcode`@=11
\def\secteqno{\@addtoreset{equation}{section}%
\def\theequation{\thesection.\arabic{equation}}}
\catcode`@=12
\secteqno
\newcommand{\be}{\begin{equation}}
\newcommand{\ee}{\end{equation}}
\newcommand{\bea}{\begin{eqnarray}}
\newcommand{\eea}{\end{eqnarray}}
\newcommand{\nn}{\nonumber}
\newcommand{\bref}[1]{(\ref{#1})}

\begin{document}
\thispagestyle{empty}
\hfill June 22, 2001 \null\par
\hfill KEK-TH-772\null\par
\hfill TOHO-FP-0169 \null\par
\vskip 20mm
\begin{center}
{\Large\bf Classical AdS Superstring Mechanics}\par
\vskip 20mm
{\large Machiko\ Hatsuda\footnote{mhatsuda@post.kek.jp} and~Kiyoshi~Kamimura\footnote{kamimura@ph.sci.toho-u.ac.jp}$^\dagger$}\par
\medskip
Theory Division,\ High Energy Accelerator Research Organization (KEK),\\
\ Tsukuba,\ Ibaraki,\ 305-0801, Japan \\
$\dagger$ Department of Physics,\ Toho University,
\ Funabashi,\ 274-8510, Japan\\
\medskip
\vskip 10mm
\end{center}
\vskip 10mm
\begin{abstract}
We analyze the anti-de Sitter (AdS) superparticle and superstring systems
described in terms of supermatrix valued coordinates
proposed by Roiban and Siegel.  
This approach gives simple symmetry transformations and
equations of motion.
We examine their $\kappa$-transformations, 
infinite reducibility and $\kappa$-gauge fixing conditions. 
A closed first class constraint set for the AdS superparticle
is $GL(4|4)$ covariant and
keeping superconformal symmetry manifestly.
For the AdS superstring $\sigma$-dependence 
breaks the $GL(4|4)$ covariance,
where supercovariant derivatives and currents satisfy the
inhomogeneous $GL(4|4)$.
A closed first class constraint set for the AdS superstring
turns out to be 
the same as the one for a superstring in flat space,
namely ${\cal ABCD}$ constraints.
\end{abstract} 
\par
\newpage
\setcounter{page}{1}
\parskip=7pt
%%%%%%%%%%%%%%%%%%%%%%%%%%%%%%%%%%%%%%%%%%%%%%%%%%%%%%%%%%%%%%%%%%%%%%%%%
\section{ Introduction}\par
\indent

The AdS/CFT correspondence conjectured by \cite{Mal}
requires appropriate descriptions of the superstring theories in anti-de Sitter (AdS) spaces corresponding to the conformal field theories (CFT).
Since AdS spaces contain the Ramond-Ramond flux, the manifest space-time supersymmetric 
Green-Schwarz formalism is more suitable than the Neveu-Schwarz-Ramond formalism.
The superstring theory in the AdS$_5\times$S$^5$ background, 
known to be dual of the four dimensional CFT,
is investigated at the classical level \cite{MeTsy,Ram,MeTlc}
and at the semi-classical level \cite{RamQ} in the Green-Schwarz formalism.
The global symmetry of the AdS$_5\times$S$^5$ space is $SU(2,2|4)$. 
The AdS superstring action is  
constructed
based on a coset ${SU(2,2|4)}/(SO(4,1)\times SO(5))$ \cite{MeTsy}.
Recently an alternative approach has been proposed by Roiban and Siegel \cite{RWS} 
based on a coset
${GL(4|4)}/{(Sp(4)\times GL(1))^2}$,
which is obtained by 
Wick rotations and introducing scaling degrees of freedom 
into ${SU(2,2|4)}/(SO(4,1)\times SO(5))$.
In this alternative approach the coordinate is supermatrix valued.
The Lobachevski metric follows from this matrix parametrization
rather than the conventional exponential parametrization.  
In this paper we follow this approach to examine the 
classical mechanics of the AdS superparticle and superstring.

When an element of the coset is given as $Z_M^{~A}\in {GL(4|4)}/{(Sp(4)\times GL(1))^2} $,
the right invariant and left invariant one-forms are constructed as $(dZZ^{-1})_M^{~N}$
and $(Z^{-1}dZ)_A^{~B}$ respectively. If we introduce its canonical conjugates, $P_A^{~M}$,
the global symmetry generators and the supercovariant derivatives can be written as
$G_M^{~N}=(ZP)_M^{~N}$ and $D_A^{~B}=(PZ)_A^{~B}$ respectively
as we will see in next section.
The global symmetry generators 
$GL(4|4)\ni G_{MN}$ satisfy
\bea
[G_{MN} ,G_{LK}]&=&
-\Omega_{LN}G_{MK}+\Omega_{MK}G_{LN}
\label{MM}~~,
\eea 
with an antisymmetric metric $\Omega$ which raises and lowers the indices.
On the other hand the supercovariant derivatives satisfy 
\bea
[D_{AB} ,D_{CD}]&=&
\Omega_{CB}D_{AD}-\Omega_{AD}D_{CB}
\label{DD}~~.
\eea 
The $D$'s correspond to $\tau$ development, so the supercovariant derivative
governs the AdS superparticle.
The covariant derivatives 
commute/anticommute with the global charges, $[G_{MN},D_{AB}]=0$.
The fermionic components of supercovariant derivatives are nothing but the fermionic constraints
whose first class part generate $\kappa$-symmetries.
For a flat superparticle supercovariant derivatives are $p_\mu$ and $d_\alpha$,
and the closed first class constraint set is given by bi-linear forms of covariant derivatives,
$p^\mu p_\mu=0$ and $p_\mu \gamma^\mu d=0$ 
\cite{WS,WS88,WSTXT}.
For a conformal particle covariant derivatives are 
conformal group elements, and 
 the closed first class constraint set is obtained by
squaring the covariant derivatives \cite{WSMH,WSMH2}.
The AdS superparticle is also the case
where a sufficient first class constraint set is 
expressed as
\bea
(D^2)_A^{~~C}=D_A^{~~B}D_B^{~~C}=0~~.\label{DD}
\eea 
They satisfy the following algebra
\bea
[{D^2}_{AB},{D^2}_{CD}]={D^3}_{AD}\Omega_{CB}-{D^3}_{CB}\Omega_{AD}+
{D^2}_{AD}D_{CB}
-{D^2}_{CB}D_{AD}~~\label{DDDD}
\eea
and are reducible,
\bea
D_A^{~~B}{D^2}_B^{~~C}-{D^2}_A^{~~B}D_B^{~~C}=0~~.\label{DDD}
\eea
The first class constraints \bref{DD} includes
the massless equation and $\kappa$-generators
representing local symmetries for the AdS superparticle.
Gauge fixing follows to the form of the first class constraints.

For an AdS superstring the $\sigma$ dependence requires 
$\sigma$ direction current $J_{AB}$ and the system is described 
in terms of $D$ and $J$ satisfying
\bea
[D_{AB} ,D_{CD}]&=&
\Omega_{CB}D_{AD}-\Omega_{AD}D_{CB}\nn\\
\left[D_{AB} ,J_{CD}\right]&=&
\Omega_{CB}J_{AD}-\Omega_{AD}J_{CB}
\label{JJ}\\
\left[J_{AB} ,J_{CD}\right]&=&0~~.\nn
\eea 
A superstring in flat space is described by 
left/right ($\partial_\tau \pm \partial_\sigma$)
chiral representations, 
but this is not the case.
The naive left/right combination $D\pm J$ does not show 
separation into two sets, but
\bea
[(D\pm J)_{AB} ,(D\pm J)_{CD}]&=&
\Omega_{CB}(D\pm 2J) _{AD}-\Omega_{AD}(D\pm 2J)_{CB}\label{DpmJ}\\
\left[(D\pm J)_{AB} ,(D\mp J)_{CD}\right]&=&
\Omega_{CB}D_{AD}-\Omega_{AD}D_{CB}\nn ~~.
\eea  
The naive replacement $D\to D\pm J$ in \bref{DD}-\bref{DDD}
does not work as the
stringy extension of the local symmetry constraint set.
To find the local symmetry constraint set
is not trivial as in the AdS superparticle case,
and one needs to start with the $\kappa$-invariant action for the AdS superstring.

It is interesting that the algebra \bref{JJ}
is recognized as an inhomogeneous $GL(4|4)$
and it is obtained from the following algebra
\bea
[D_{AB} ,D_{CD}]&=&
\Omega_{CB}D_{AD}-\Omega_{AD}D_{CB}\nn\\
\left[D_{AB} ,J_{CD}\right]&=&
\Omega_{CB}J_{AD}-\Omega_{AD}J_{CB}
\label{JJR}\\
\left[J_{AB} ,J_{CD}\right]&=&
\Omega_{CB}D_{AD}-\Omega_{AD}D_{CB}\nn
~~
\eea
 by the In$\ddot{\rm o}$n$\ddot{\rm u}$-Wigner contraction \cite{IW}
with $J\to RJ$ and $R\to \infty$ limit.
In contrast to \bref{DpmJ} the left/right chiral separation is possible for $R=1$,
\bea
[(D\pm J)_{AB} ,(D\pm J)_{CD}]&=&
\Omega_{CB}\{(1+\frac{1}{R^2})D\pm 2J\} _{AD}
-\Omega_{AD}\{(1+\frac{1}{R^2})D\pm 2J\}_{CB}\nn\\
\label{DpmJe}\\
\left[(D\pm J)_{AB} ,(D\mp J)_{CD}\right]&=&
\Omega_{CB}(1-\frac{1}{R^2})D_{AD}
-\Omega_{AD}(1-\frac{1}{R^2})D_{CB}\nn ~~.
\eea
The AdS superstring case \bref{DpmJ} is given for a case with $R\to \infty$.
In this case rescaling ambiguity of $J$ is remained.
We will construct the left/right chiral algebra up to constraints
in the section 4 where chiral combinations are
 $D_{\rm bose}\pm J_{\rm bose}$ and $D_{\rm fermi}\pm (1/2)J_{\rm fermi}$.

A local symmetry constraint set for the AdS superstring 
is constructed and compared with the one for a superstring in a flat space, 
namely ${\cal ABCD}$ constraints \cite{WS,WS88,WSTXT}.
 A flat superstring system can be described in
 two approaches.
 One is the usual canonical approach that includes  
 both first class and second class constraints,
 and products of two constraints can be set to be zero consistently \cite{HrKm}.
 Another one is the first class approach that includes only
 first class constraints and square of the 
 supercovariant derivatives are taken as first class constraints \cite{WS}.
It is shown that
these approaches, \cite{HrKm} and \cite{WS},
give the same physical states
by analyzing in the light-cone gauge 
 \cite{KT}.
An advantage of this first class approach is to avoid complicated second class constraints.
Disadvantage is that the first class constraints are infinitely reducible, 
so practical covariant quantum computation is difficult
\cite{ABCD}. 
In this paper we will focus on the classical mechanics for AdS objects
to clarify common features and different features from the ones in the flat background.

The organization of this paper is as follows.
In section 2, we perform the canonical analysis of the AdS superparticle system.
The $\kappa$ symmetry constraints and its reducibility are clarified.
The equation of motion is also given, and it has simple free form.
In section 3, we confirm the necessary conditions of the closed first class constraint set 
by direct computation from an action.
The $(D^2)_A^{~B}=0$ constraint set is analyzed in ``light-cone-like" gauge 
that suppresses half of fermionic degrees of freedom,
and then we show that $(D^2)_A^{~B}=0$ constraints reduce into GL(2)$\times$GL(4)$\cong$U(2)$\times$U(4) 
generators which turn out to represent the second class constraints.
In other words, the AdS superparticle with first class constraint set, $(D^2)_A^{~B}=0$,
is equal to the system with the first and second class constraints.
In section 4, the AdS superstring system is examined in analogous to the AdS superparticle case.
The $\kappa$-gauge fixing and equations of motions are examined, and 
we show that the complex gauge proposed in \cite{RWS} can not be applied.
The closed first class constraint set is shown to correspond to the one in a flat background, namely ${\cal ABCD}$ constraints.
\par
%%%%%%%%%%%%%%%%%%%%%%%%%%%%%%%%%%%%%%%%%%%%%%%%%%%%
\section{ Classical AdS superparticle mechanics}\par

A supersymmetric object propagating in AdS$_5\times$S$^5$ space is 
described using  by the following coset superspace
and the coset elements are parameterized including quotient parameters as \cite{RWS}
\bea
\frac{GL(4|4)}{(Sp(4)\times GL(1))^2}\ni  Z_M^{\ \ A}~~,~~
Z_M^{\ \ A}=\pmatrix{Z_m^{\ \ a}&Z_{m}^{\ \ \bar{a}}\cr Z_{\bar{m}}^{\ \ a}&Z_{\bar{m}}^{\ \ \bar{a}} }
=\pmatrix{X_m^{\ \ a}&\Theta_{m}^{\ \ \bar{a}}\cr \bar{\Theta}_{\bar{m}}^{\ \ a}&\bar{X}_{\bar{m}}^{\ \ \bar{a}}},
\eea  
where $M$ and $A$ are transformed by global $GL(4|4)$ and local 
$(Sp(4)\times GL(1))^2$ corresponding to 
the global coordinate transformations and the local Lorentz and the dilatation respectively.  
Throughout this paper we put full components of parameters $8^2=64$ as $Z_M^{~~A}$,
then gauge degrees of freedom are eliminated by constraints.
The left invariant current is 
\bea
J_A^{\ \ B}&=&Z_{~~A}^{-1~ M}dZ_M^{\ \ B}~~.\label{JJJJ}
\eea
A $GL(4)$ matrix $M_{ab}$ 
is decomposed into a trace part, symmetric part and traceless-antisymmetric part
\bea
M_{ab}=-\frac{1}{4}\Omega_{ab}M^c_{\ \ c}+M_{(ab)}+M_{\langle ab\rangle}=-\frac{1}{4}\Omega~{\rm tr}M+(M)+\langle M\rangle,
\eea
where $\Omega$ is antisymmetric $SP(4)$ metric raising and lowering indices in the NW rule.
The coset parts of the left invariant currents are 
$J_{\langle ab\rangle}$,  $J_{\langle \bar{a}\bar{b}\rangle}$, $J_{a\bar{b}}$ and $J_{{a}\bar{b}}$,
and local Lorentz $Sp(4)$, dilatation $GL(1)$ connections are
$J_{(ab)}$, tr$J$ respectively.
\par
%%%%%%%%%%%%%%%%%%%%%%%%%%%%%%%%%%%%%%%%%%%%%%%%%%%%%%%%%%%%%%%%%%%%%%%%%%%%%%%%%%
%\subsection{Canonical analysis of an AdS superparticle}\par

The action for a superparticle is given in terms of the 1-form currents $J_{AB}=d\sigma^\mu J_{\mu~AB}$;
\bea
S&=&-\int d\tau \frac{1}{2e}\left(
J_0^{\langle ab\rangle}J_{0~\langle ab\rangle}
-J_0^{\langle \bar{a}\bar{b}\rangle}J_{0~\langle \bar{a}\bar{b}\rangle}
\right)=\int d\tau {\cal L}~~.\label{adsSUPA}
\eea
Canonical momentum for $Z_M^{~~A}$ is defined as
\bea
P_A^{\ \ M}&=&\frac{\delta^r S}{\delta \dot{Z}_M^{\ \ A}}(-)^A=
\pmatrix{P_a^{~~m}&\bar{\zeta}_a^{~~\bar{m}}\cr \zeta_{\bar{a}}^{~~m}
&\bar{P}_{\bar{a}}^{~~\bar{m}}}\label{ppp},
\eea
and the Hamiltonian is given by
\bea
{\cal H}=[\sum P_A^{\ \ M}\dot{Z}_M^{\ \ A}(-)^A -{\cal L}]~~.\label{Hamil}
\eea
Corresponding to \bref{ppp} we choose the canonical poisson bracket as
\bea
[ Z_M^{~~A},P_{B}^{~~N}]=(-1)^{A}\delta_B^A\delta_M^N
\eea
in order to preserve supertrace
\bea
&&{\rm STr} fg =\sum f^{AM}g_{MA}(-)^A~~,\nn\\
&&[ {\rm STr} (fZ), {\rm STr} (Pg)]=-{\rm STr} (fg).
\eea

The AdS superparticle action \bref{adsSUPA} is rewritten as
\bea
S&=&-\int d\tau\frac{1}{2e}
{\rm tr}\left(
\langle ~X^{-1}v(\dot{X}-\Theta\bar{X}^{-1}\dot{\bar{\Theta}})~\rangle
\langle~X^{-1}v(\dot{X}-\Theta\bar{X}^{-1}\dot{\bar{\Theta}})~\rangle \right.\nn\\
&&~~~~~~~~~~~~-\left.
\langle ~\bar{X}^{-1}\bar{v}(\dot{\bar{X}}-\bar{\Theta}{X}^{-1}\dot{{\Theta}})~\rangle
\langle ~\bar{X}^{-1}\bar{v}(\dot{\bar{X}}-\bar{\Theta}{X}^{-1}\dot{{\Theta}})~\rangle
\right),\label{adsSUPA1}\\
&&v=(I-\Theta\bar{X}^{-1}\bar{\Theta}X^{-1})^{-1}\nn\\
&&\bar{v}=(I-\bar{\Theta}{X}^{-1}{\Theta}\bar{X}^{-1})^{-1}\nn~~.
\eea
Canonical conjugates are
\bea
P_a^{~~m}&\equiv&\frac{\delta S}{\delta \dot{X}_m^{~~a}}=
-\frac{1}{e}X^{-1}v(\dot{X}-\Theta\bar{X}^{-1}\dot{\bar{\Theta}})X^{-1}v\nn\\
\bar{P}_{\bar{a}}^{~~\bar{m}}&\equiv&
-\frac{\delta S}{\delta \dot{\bar{X}}_{\bar{m}}^{~~\bar{a}}}=
-\frac{1}{e}\bar{X}^{-1}\bar{v}(\dot{\bar{X}}-\bar{\Theta}{X}^{-1}\dot{{\Theta}})\bar{X}^{-1}\bar{v}
\nn\\
\zeta_{\bar{a}}^{~~m}&\equiv&
-\frac{\delta^r S}{\delta \dot{\Theta}_m^{~~\bar{a}}}=
-\bar{P}\bar{\Theta}X^{-1}\label{pF}\\
\bar{\zeta}_{{a}}^{~~\bar{m}}&\equiv&
\frac{\delta^r S}{\delta \dot{\bar{\Theta}}_{\bar{m}}^{~~{a}}}=
-{P}{\Theta}\bar{X}^{-1}\nn\label{pFb}.
\eea

The supercovariant derivatives are
\bea
&&D_A^{~~B}=P_A^{~~M}Z_M^{~~B}=
\pmatrix{
{\bf D}_a^{~~b}&  D_a^{~~\bar{b}}  \cr 
  \bar{D}_{\bar{a}}^{~~b} &  \bar{\bf D}_{\bar{a}}^{~~\bar{b}}
},\label{covder}\\
&&~~~~\begin{array}{cclcccl}
{\bf D}&\equiv&PX+\bar{\zeta}\bar{\Theta}&,&
\bar{\bf D}&\equiv&\bar{P}\bar{X}+{\zeta}{\Theta}\nn\\
D&\equiv&{P}{\Theta}+\bar{\zeta}\bar{X}&,&
\bar{D}&\equiv&\bar{P}\bar{\Theta}+\zeta X
\end{array}\nn
\eea
generating right $GL(4|4)$ transformations
\bea
\delta_\Lambda Z_M^{~~A}=[Z_M^{~~A},{\rm STr}D\Lambda]=
(Z%_M^{~~B}
\Lambda%_B^{~~A}.
)_M^{~~A}~~.
\eea
Its subgroups $GL(1),Sp(4),\overline{GL(1)},\overline{Sp(4)}$ 
are local gauge symmetries, and 
their generators are set to be constraints in our approach
\bea
{\rm tr}{\bf D}=({\bf D})= {\rm tr}\bar{\bf D}=(\bar{\bf D})=0~~.\label{sp4gl1}
\eea

The last two fermionic equations of \bref{pF} are primary constraints,
\bea
D_a^{~~\bar{b}}=0,~~\bar{D}_{\bar{a}}^{~~b}=0~~\label{1st2nd}
\eea
satisfying following poisson brackets 
\bea
[D_a^{~~\bar{b}} ,\bar{D}_{\bar{c}}^{~~d}]&=&
\delta_a^{d}\bar{\bf D}_{\bar{c}}^{~~\bar{b}}-
\delta_{\bar{c}}^{\bar{b}}{\bf D}_{a}^{~~{d}}
\label{FF}~~,
\left[D,D\right]=[\bar{D},\bar{D}]~=~0~~.
\eea 
The Hamiltonian \bref{Hamil} is obtained as
\bea
{\cal H}&=&-eA_{\rm P}+{\rm tr}[D\bar{\lambda}
+\bar{D}\lambda]\label{HamilSUPA}\\
A_{\rm P}&=&\frac{1}{2}{\rm tr}[\langle {\bf D}\rangle ^2 -\langle \bar{\bf D}\rangle ^2
]=0~~\label{ASUPA}
\eea
where $A_{\rm P}=0$ is a secondary constraint for $\dot{\Pi}_e=0$
and multipliers determined from consistency as 
$\lambda=\bar{\lambda}=0$ .
Because of the bosonic constraint \bref{ASUPA}
the rank of the right hand side of \bref{FF} is half the number of its maximal rank,
and whose zero modes are given as 
\bea
\left(
\delta_a^{d}\bar{\bf D}_{\bar{c}}^{~~\bar{b}}-
\delta_{\bar{c}}^{\bar{b}}{\bf D}_{a}^{~~{d}}
\right)
\left(
\delta_d^{e}\bar{\bf D}_{\bar{b}}^{~~\bar{f}}+
\delta_{\bar{b}}^{\bar{f}}{\bf D}_{d}^{~~{e}}
\right)
=\frac{1}{2}\delta_a^e\delta_{\bar{c}}^{\bar{f}}A_{\rm P}\approx 0
%+({\rm tr}{\bf D},({\bf D}), {\rm tr}\bar{\bf D},(\bar{\bf D}))
\label{zeromode}
,
\eea
where equality holds up to gauge constraints \bref{sp4gl1}.
Half of the fermionic constraints are first class and another half are 
second class constraints as in a flat superparticle.
The fermionic constraints  $D$ and $\bar{D}$ in \bref{1st2nd}
are projected into first class constraints
using $\langle{\bf D}\rangle$ and $\langle\bar{\bf D}\rangle$ as
\bea
{{\cal B}_{\rm P}}_a^{~~\bar{b}}&=&\langle {\bf D}\rangle_a^{~~{b}} D_b^{~~\bar{b}}  
+D_a^{~~\bar{a}}\langle \bar{\bf D}\rangle_{\bar{a}}^{~~\bar{b}}\nn\\
  \bar{\cal B}_{{\rm P}~\bar{a}}^{~~~~{b}}&=&\langle \bar{\bf D}\rangle_{\bar{a}}^{~~\bar{b}} \bar{D}_{\bar{b}}^{~~{b}} +\bar{D}_{\bar{a}}^{~~{a}}\langle {\bf D}\rangle_a^{~~b}\label{1stBB}~~.
\eea 
They generate following $\kappa$-transformations 
\bea
\delta_{\kappa,\bar{\kappa}}Z&=&[Z,{\rm tr}(\bar{\cal B}\kappa-{\cal B}\bar{\kappa})]\label{kappatr}\\
\delta_{\kappa,\bar{\kappa}}X&=&%X(\langle D\bar{\kappa}\rangle +\langle \kappa\bar{D}\rangle)+
\Theta(\bar{\kappa}\langle {\bf D}\rangle+\langle \bar{\bf D}\rangle \bar{\kappa})\nn\\
\delta_{\kappa,\bar{\kappa}}\bar{X}&=&%\bar{X}(\langle \bar{\kappa}D\rangle +\langle \bar{D}\kappa\rangle)+
\bar{\Theta}({\kappa}\langle \bar{\bf D}\rangle+\langle {\bf D}\rangle {\kappa})\nn\\
\delta_{\kappa,\bar{\kappa}}\Theta&=&%\Theta(\langle \bar{\kappa}D\rangle +\langle \bar{D}\kappa\rangle)+
X({\kappa}\langle \bar{\bf D}\rangle+\langle {\bf D}\rangle {\kappa})\nn\\
\delta_{\kappa,\bar{\kappa}}\bar{\Theta}&=&%\bar{\Theta}(\langle D\bar{\kappa}\rangle +\langle \kappa\bar{D}\rangle)%+
\bar{X}(\bar{\kappa}\langle {\bf D}\rangle+\langle \bar{\bf D}\rangle \bar{\kappa})\nn~~.
\eea
Both parameters $\kappa$ and $\bar{\kappa}$  have following zero modes,
and this reducibility continues infinitely
\bea
\delta\kappa&=&\langle {\bf D}\rangle {\kappa}_1~-{\kappa}_1\langle \bar{\bf D}\rangle~~,~~
\delta\kappa_1=\langle {\bf D}\rangle {\kappa}_2+{\kappa}_2\langle \bar{\bf D}\rangle~~,~~
\cdots~~,
\label{k0k1k2}\\
\delta\bar{\kappa}&=&\langle \bar{\bf D}\rangle \bar{\kappa}_1~-\bar{\kappa}_1\langle {\bf D}\rangle~~,~~
\delta\bar{\kappa}_1=\langle \bar{\bf D}\rangle \bar{\kappa}_2+\bar{\kappa}_2\langle {\bf D}\rangle~~,~~
\cdots\nn~~,
\eea
where $A_{\rm P}=0$ is used.
This fact leads to that only each half components of $\Theta$ and $\bar{\Theta}$ can be 
fixed by using $\kappa$-symmetries.
Therefore the complex gauge proposed in the literature \cite{RWS} can not be imposed
for the AdS superparticle.

Classical equations of motion are given by this Hamiltonian with for example $e=1$
\bea
\dot{Z}_M^{~~A}&=&\left[{Z}_M^{~~A},{\cal H}\right]
={Z}_M^{~~B}\Gamma_{{\rm P}~B}^{~~~~A}~~,~~
\Gamma_{{\rm P}~}=
\pmatrix{\langle {\bf D}\rangle&0\cr
0&\langle \bar{\bf D} \rangle } \label{eqs}
\eea
where the fermionic constraints \bref{1st2nd} is used. 
The second derivative of $Z$ leads to 
\bea
\ddot{Z}-\dot{Z}\Gamma_{\rm P}=0~~,\label{conn}
\eea 
and it suggests that $\Gamma_{\rm P}$ is interpreted as the induced connection onto the world line.
$\langle {\bf D}\rangle$ and  $\langle \bar{\bf D}\rangle$,
are conserved ``momenta" of this system.
 They satisfy 
\bea
[\langle {\bf D}\rangle_{\langle ab\rangle},
\langle {\bf D}\rangle_{\langle cd\rangle}]=
\Omega_{\langle c|\langle b}({\bf D})_{a\rangle |d\rangle}-
\Omega_{\langle a|\langle d}({\bf D})_{c\rangle |b\rangle}~~
\eea
which is the AdS momentum algebra, and the similar form for bared (S$^5$) sector.
\bref{eqs} represents a free particle.

Solutions of \bref{eqs} are
\bea
{Z}(\tau)&=&{Z}_0 e^{\Gamma_{\rm P}\tau}
\label{sol}~~.
\eea
The world line element is calculated as
\bea
ds^2&=&STr ~J_0^{~~2}\mid_{\rm bosonic~part}=STr~(Z^{-1}\dot{Z})^2=STr~\Gamma_{\rm P}^{~~2}=\frac{1}{2}A_{\rm P}=0
\label{nwl}~~
\eea 
where the constraints \bref{ASUPA} is used.
Therefore the free superparticle moves along the null geodesic of the AdS(S) space.

\par
%%%%%%%%%%%%%%%%%%%%%%%%%%%%%%%%%%%%%%%%%%%%%%%%%%%%%%%%%%%%%%%%%%%%%%%%%%%%%%%%%%
\section{Local symmetries of the AdS superparticle}\par

Local symmetries are expressed by a closed first class constraint set.
In the previous section we perform a canonical analysis of a superparticle 
where products of constraints
including second class are set to be zero consistently.
In this section we examine an alternative approach that contains only first class constraints
without second class constraints. 
We obtain a closed first class constraint set and then examine the physical states
in a ``light-cone-like" gauge comparing with the one described by the previous approach.

Although we know that \bref{DD} is a sufficient closed first class constraint set,
we calculate it directly from $A_P$ in \bref{ASUPA} and $D,\bar{D}$ in \bref{1st2nd},
then obtain a necessary first class constraint set as
\bea
{\cal A}_{\rm P}&=&\frac{1}{2}{\rm tr}\left( \langle {\bf D}\rangle ^2 -\langle \bar{\bf D}\rangle ^2
+2D\bar{D}
\right)\nn\\
{{\cal B}_{\rm P}}_a^{~~\bar{b}}&=&\langle {\bf D}\rangle_a^{~~{b}} D_b^{~~\bar{b}}  
+D_a^{~~\bar{a}}\langle \bar{\bf D}\rangle_{\bar{a}}^{~~\bar{b}}\nn\\
  \bar{\cal B}_{{\rm P}~\bar{a}}^{~~{b}}&=&\langle \bar{\bf D}\rangle_{\bar{a}}^{~~\bar{b}} \bar{D}_{\bar{b}}^{~~{b}} +\bar{D}_{\bar{a}}^{~~{a}}\langle {\bf D}\rangle_a^{~~b}\label{SUPAff}\\
{c_{\rm P}}_{(ab)}&=&D_{(a|}^{~~\bar{a}}\bar{D}_{\bar{a}|b)}~~~
,~~~\bar{c}_{{\rm P}~(\bar{a}\bar{b})}=\bar{D}_{(\bar{a}|}^{~~a}{D}_{a|\bar{b})}\nn\\
c_{{\rm P}~\langle ab\rangle}&=& D_{\langle a|}^{~~\bar{a}}\bar{D}_{\bar{a}|{b}\rangle}~~~,
~~~\bar{c}_{{\rm P}~\langle \bar{a}\bar{b}\rangle}= \bar{D}_{\langle \bar{a}|}^{~~{a}}{D}_{a|\bar{b}\rangle}\nn
~~.
\eea
The reparametrization constraint is ${\cal A}_{\rm P}$
which is the Hamiltonian \bref{Hamil} to preserve
all symmetry constraints up to bi-linears of constraints.
The $\kappa$-symmetry constraints are
${\cal B}_{\rm P}$ and $\bar{\cal B}_{\rm P}$, and others are required for closure.
The first class constraint set \bref{SUPAff} can be written as
\bea
{D^2}_A^{~~B}-\frac{1}{4+4}\delta_A^B~{\rm Tr} D^2=0~~,\label{32}
\eea 
where Tr$M=\sum M^A_{~~A}$ should not be confused with STr$M=\sum (-)^AM^A_{~~A}$.
The number of constraints for ${\cal A}_{\rm P}$, ${\cal B}_{\rm P}$, 
$\bar{\cal B}_{\rm P}$ are $1$, $16$, $16$ respectively.
 Total number of constraints is $1+16\times 2 +10\times 2+5\times 2=63=(4+4)^2-1$
 corresponding to the dimension of $GL(4|4)/GL(1)$.
This result beginning with the AdS superparticle action \bref{adsSUPA} 
differs from \bref{DD} just trace part of $D^2$.
Imposing both trace part and supertrace part of $D^2$ constraints
means massless constraints in both AdS$_5$ space and S$^5$ space,
which is too strong for single particle.

\par
%%%%%%%%%%%%%%%%%%%%%%%%%%%%%%%%%%%%%%%%%%%%%%%%%%%%%%%%%%%%%%%%%%%%%%%%%%%%%%%%%%
%\subsection{$\kappa$-gauge fixing of an AdS superparticle}\par

The constraints are reduced in a light-cone-like gauge 
as follows.
Using with following projection operators of AdS spinors 
\bea
{{\cal P}_\pm}_a^{~~b}=\frac{1}{2}(\gamma_\mp \gamma_\pm)_a^{~~b}~~,~~
{\cal P}_+=
\pmatrix{{\bf 1}&0\cr 0&0}~~,~~
{\cal P}_-=
\pmatrix{0&0\cr 0&{\bf 1}}
 \label{Ppm}~~,
\eea
supercovariant derivatives are decomposed into
\bea
D_a^{~~\bar{b}}&=&{\cal P}_+D_+ +{\cal P}_-D_-~~~,~~~
\bar{D}_{\bar{a}}^{~~{b}}=\bar{D}_+{\cal P}_+ +\bar{D}_-{\cal P}_-~~,~~\label{proj}\\
\langle {\bf D}\rangle_a^{~~b}&=&
{\cal P}_+\langle {\bf D}\rangle_+ {\cal P}_-
+{\cal P}_-\langle {\bf D}\rangle_- {\cal P}_+
+{\cal P}_+\langle {\bf D}\rangle_{\perp }{\cal P}_+
+{\cal P}_-\langle {\bf D}\rangle_{\perp }{\cal P}_-
 ~~.\nn
\eea 
Assuming ${\cal P}_+\langle {\bf D}\rangle_+ {\cal P}_-
\neq 0$, ${\cal B}_{\rm P}=0=\bar{\cal B}_{\rm P}$ are solved for 
$D_-$ and $\bar{D}_+$.
This allows the following light-cone-like gauge
\bea
&&{\rm tr}[\langle {\bf D}\rangle_+ \gamma_+]%\equiv
%{\rm tr}[{\cal P}_+\langle {\bf D}\rangle_+ {\cal P}_-\gamma_+]
=4p_+ \nn~~,
\\
&&\bar{\Theta}_{-\bar{m}}^{~~~a}\equiv \bar{\Theta}_{\bar{m}}^{~~b}{\cal P}_{-b}^{~~a}=0~~~,~~~\label{lcg}\\
&&\Theta_{+m}^{~~~\bar{c}}\equiv
X_m^{~~a}{\cal P}_{+a}^{~~b}{X^{-1}}_{b}^{~~n}\Theta_{n}^{~~\bar{c}} =
0~~,\nn
\eea  
where $X{\cal P}_\pm X^{-1}$ matrix can be used as the projection for $\Theta_m^{~~\bar{a}}$
because of invertibility of $X_m^{~~a}$. 
The $p_+$ is assumed not to be zero.% but may not be constant in general.
These gauge fixing conditions \bref{lcg} make ${\cal B}_{{\rm P}+}$
 and $\bar{\cal B}_{{\rm P}-}$ 
to be second class 
and they are solved as
\bea
\bar{\zeta}_-&=&-\frac{1}{2p_+}[\langle {\bf D}\rangle_\perp D_++D_+\langle \bar{\bf D}\rangle]\bar{X}^{-1}
-\langle{\bf D}_{\perp}\rangle X^{-1}\Theta_-\bar{X}^{-1}
\label{zetans}\\
{\zeta}_+&=&-\frac{1}{2p_+}[\bar{D}_
-\langle {\bf D}\rangle_\perp +\langle \bar{\bf D}\rangle \bar{D}_-]\gamma_+
{X}^{-1}-\bar{P}\bar{\Theta}_+X^{-1}
\nn
\eea
where $\bar{\zeta}_-$ and $\zeta_+$ are conjugates of $\bar{\Theta}_-$ and $\Theta_+$
 defined through ${\cal P}_-\bar{\zeta}$ 
and $\zeta X{\cal P}_{+}X^{-1~~}$ respectively.
${\cal A}_{\rm P}=0$ can be written as
\bea
{\rm tr}[\langle {\bf D}\rangle_- \gamma_-]&=&\frac{-1}{2p_+}{\rm tr}
[\langle {\bf D}\rangle_\perp^2-\langle \bar{\bf D}\rangle^2]
+\frac{1}{p_+^{~~2}}{\rm tr}[
\langle {\bf D}\rangle_{\perp }\gamma_+D_+\bar{D}_-
-\langle \bar{\bf D}\rangle \bar{D}_-\gamma_+D_+]\label{alc}~~.
\eea

Remaining components of constraints ${\cal B}_{{\rm P}-}$, $\bar{\cal B}_{{\rm P}+}$, 
$c_P$ and $\bar{c}_{\rm P}$
 are first class constraints
written by $D_+$, $\bar{D}_-$, $p_+$, $\langle {\bf D}\rangle_\perp$ and
$\langle \bar{\bf D}\rangle$ 
\bea
{\cal B}_{{\rm P}-}&=&\frac{1}{4p_+^2}\gamma_+D_
+{\rm tr}[{c_{\rm P}}_{a_+~b_-}\langle {\bf D}\rangle_\perp\gamma_+-
\langle \bar{\bf D}\rangle \bar{D}_-\gamma_+D_+]=0\nn~~,\\
\bar{\cal B}_{{\rm P}+}&=&\frac{1}{4p_+^2}\bar{D}_-\gamma_+
{\rm tr}[{c_{\rm P}}_{a_+~b_-}\langle {\bf D}\rangle_\perp\gamma_+-
\langle \bar{\bf D}\rangle \bar{D}_-\gamma_+D_+]=0\nn~~,\nn\\
{c}_{\rm P}&=&D_{+}\bar{D}_-~~,
\nn\\
\bar{c}_{\rm P}&=&-\frac{1}{p_+}\langle \bar{\bf D}\rangle  \bar{D}_-\gamma_+D_+\nn~~.
\eea 
It is noted that ${\cal B}_{\rm P}$ and $\bar{\cal B}_{\rm P}$ are fixed 
up to a square of fermionic constraints which 
can be set to zero in the second class approach,
so ${\cal B}_{{\rm P}-}$ and $\bar{\cal B}_{{\rm P}+}$ do not produce 
further constraints.
But this is not the case in the first class approach.
We take a sufficient closed constraint set to examine 
the restricted states,
\bea
c_{\rm P}\to~D_+\bar{D}_-=0,~~ {\cal B},\bar{\cal B},\bar{c}_{\rm P}\to~\bar{D}_-\gamma_+D_+
%\mid_{\rm traceless}
=0~~.\label{cpbbc}
\eea 
Independent supercovariant derivatives $D_+$ and $\bar{D}_-$ satisfy 
\bea
[(D_+)_{a}^{~~\bar{b}},(\bar{D}_-)_{\bar{c}}^{~~d}]&=&-p_+\delta_{\bar{c}}^{\bar{b}}(\gamma_-)_{a}^{~~d}~~.
\label{ddcr}
\eea
Independent fermionic degrees of freedom are 8 $Q_+$ and 8 $\bar{Q}_-$ 
\bea
Q_+=X{\cal P}_+X^{-1}Q~~&,&~~
Q_m^{~~\bar{n}}=X\bar{\zeta}+\Theta\bar{P}
\label{qqcv}\\
\bar{Q}_-=\bar{Q}X{\cal P}_-X^{-1}~~&,&~~
\bar{Q}_{\bar{m}}^{~~n}=\bar{X}{\zeta}+\bar{\Theta}{P}
\nn
\eea
which satisfy \bref{MM} and anticommute with $D_+$ and $\bar{D}_-$.
%and they are dynamical degrees of freedom.
The poisson bracket \bref{ddcr} are rewritten as the four sets of 4-dimensional 
Clifford algebras, and further can be written by two sets of 4-dimensional
creation-annihilation commutators in U(2)$\times$ U(4) basis as
\bea
\left[\bar{\bf d}_i^{\bar{a}},{\bf d}_j^{\bar{b}}\right]=\delta_{ij}\delta^{\bar{a}\bar{b}}~~.
\eea
In \bref{cpbbc}
 $D_+\bar{D}_-$'s are generators of GL(2)$\cong$U(2)
and $\bar{D}_-\gamma_+D_+$'s are generators of GL(4)$\cong$U(4),
which are imposed as
\bea
\langle {\bf d}\mid V \mid {\bf d}\rangle=0~~,~~
V=\{D_+\bar{D}_-~~,~~\bar{D}_-\gamma_+D_+%\mid_{\rm traceless}
\}
\label{gb}
\eea
where 
\bea
\mid {\bf d}\rangle &=&\{
 \mid \phi\rangle, ~{\bf d}\mid \phi\rangle,~ \cdots,~{\bf d}^8\mid \phi\rangle
  \}\label{states}~~
 \eea
with the ground state defined by $ \bar{\bf d}\mid \phi\rangle=0$.
The conditions \bref{gb} is realized on the states $\mid {\bf d}\rangle$ as
\bea
&&(V_0-v_0)\mid {\bf d}\rangle=0~~,~~V_-\mid {\bf d}\rangle=0
\label{vdvd}\\
&&
V_0=\{V_{0i}={\bf d}^{\bar{a}}_i\bar{\bf d}^{\bar{a}}_i|_{\bar{a}:{\rm sum}, i:{\rm non}\mbox{-}{\rm sum}},~~
V_0^{~\bar{a}}={\bf d}^{\bar{a}}_i\bar{\bf d}^{\bar{a}}_i|_{\bar{a}:{\rm non}\mbox{-}{\rm sum}, i:{\rm sum}}\}\nn\\
&&V_-=\{V_{-ij}={\bf d}^{\bar{a}}_i\bar{\bf d}^{\bar{a}}_j|_{\bar{a}:{\rm sum}, i>j},~~
V_-^{~\bar{a}\bar{b}}={\bf d}^{\bar{a}}_i\bar{\bf d}^{\bar{b}}_i|_{\bar{a}>\bar{b}, i:{\rm sum}}\}\nn~~,
\eea
and restricted states are 
$\mid \phi\rangle$ for $v_0=0$ or ${\bf d}_1^{\bar{1}}\mid \phi\rangle$ for $(v_0)_1=(v_0)^{\bar{1}}=1$
as highest weight states depending on the $v_0$.
These are the same higest weight states with the one for a flat case. 
It can be obtained by imposing GL(2)$\times$GL(4) invariance
while it is obtained by SO(8) invariance for a flat case.
In other words this closed first class constraint set $(D^2)_A^{~B}=0$ 
is conformal extension of the one for a  
flat superparticle \cite{KT}.
\par
%%%%%%%%%%%%%%%%%%%%%%%%%%%%%%%%%%%%%%%%%%%%%%%%%%%%
\section{ Classical AdS superstring mechanics}\par

We begin with the action for a superstring given by \cite{RWS}
\bea
S&=&\int d\sigma^2~ \frac{1}{2}\left\{
-\sqrt{-g}g^{\mu\nu}(J^{\langle ab\rangle}_\mu J_{\langle ab\rangle \nu}
-J^{\langle \bar{a}\bar{b}\rangle}_\mu J_{\nu~\langle \bar{a}\bar{b}\rangle })
 + \frac{k}{2}\epsilon^{\mu\nu}
 (E^{1/2}J^{ a\bar{b}}_\mu J_{\nu a\bar{b}~}
-E^{-1/2}J^{\bar{a}{b}}_\mu J_{\nu ~ \bar{a}{b}~})
\right\}\nn\\
&&\label{stac}
\eea
with $k=\pm 1$.
Canonical conjugates  $P_A^{~~M}$ are defined in \bref{ppp}, and we also introduce useful
variables 
\bea
\tilde{P}_a^{~~m}&=&P_a^{~~m}
+\frac{k}{2} E^{-1/2}(\bar{X}^{-1}\bar{v}\bar{\Theta}X^{-1})^{\bar{b}m}
(\bar{J}_1)_{\bar{b}a}
=-\sqrt{-g}(g^{00}{\bf J}_0+g^{01}{\bf J}_1)_a^{~~b}(X^{-1}v)_b^{~~m}\nn\\
\tilde{\bar{P}}_{\bar{a}}^{~~\bar{m}}&=&\bar{P}_{\bar{a}}^{~~\bar{m}}
+\frac{k}{2} E^{1/2}({X}^{-1}{v}{\Theta}\bar{X}^{-1})^{{b}\bar{m}}
({J}_1)_{{b}\bar{a}}
=-\sqrt{-g}(g^{00}\bar{\bf J}_0+g^{01}\bar{\bf J}_1)_{\bar{a}}^{~~\bar{b}}
(\bar{X}^{-1}\bar{v})_{\bar{b}}^{~~\bar{m}}\nn\\
\zeta_{\bar{a}}^{~~m}&=&-(\bar{P}\bar{\Theta}X^{-1})_{\bar{a}}^{~~m}
+\frac{k}{2} E^{1/2}(X^{-1})^{bm}(J_1)_{b\bar{a}}
\label{pFst}\\
\bar{\zeta}_{{a}}^{~~\bar{m}}&=&-({P}{\Theta}\bar{X}^{-1})_{{a}}^{~~\bar{m}}
+\frac{k}{2} E^{-1/2}(\bar{X}^{-1})^{\bar{b}\bar{m}}(\bar{J}_1)_{\bar{b}{a}}
\nn\label{pFbst}.
\eea
The last two fermionic equations are primary constraints corresponding to the one for the AdS superparticle \bref{1st2nd},
\bea
F_a^{~~\bar{b}}&=&E^{1/4}D_a^{~~\bar{b}}+
 \frac{k}{2}E^{-1/4}(\bar{J}_1)^{\bar{b}}_{~a}=0\nn\\
\bar{F}_{\bar{a}}^{~~{b}}&=&E^{-1/4}\bar{D}_{\bar{a}}^{~~{b}}+
\frac{k}{2} E^{1/4}({J}_1)^{b}_{~\bar{a}}=0\label{FFst}~~.
\eea
The Hamiltonian is given by
\bea
{\cal H}&=&-\frac{2}{\sqrt{-g}g^{00}}A_{\perp}
-\frac{2g^{01}}{g^{00}}A_{\parallel}+{\rm tr}[F\bar{\lambda}
+\bar{F}\lambda]\label{HamilSUST}\\
A_{\perp}&=&\frac{1}{2}{\rm tr}[\langle {\bf D}\rangle ^2 +\langle {\bf J}\rangle ^2 
-\langle \bar{\bf D}\rangle ^2-\langle \bar{\bf J}\rangle ^2 
]=0~~\label{AplSUST}\\
A_{\parallel}&=&{\rm tr}[\langle {\bf D}\rangle \langle {\bf J}\rangle  
-\langle \bar{\bf D}\rangle \langle \bar{\bf J}\rangle 
]=0~~\label{AppSUST}
\eea
with multipliers $\lambda$'s determined consistently
\bea
\lambda&=&-\frac{2}{\sqrt{-g}g^{00}}(E^{-1/4}\bar{J})
-\frac{2g^{01}}{g^{00}}(-E^{1/4}{J})\nn\\
\bar{\lambda}&=&-\frac{2}{\sqrt{-g}g^{00}}(-E^{1/4}J)-\frac{2g^{01}}{g^{00}}(E^{-1/4}\bar{J})\label{lmbd}~~,
\eea
where we denote $J_1$ as $J$ from now on.

The poisson brackets of the fermionic constraints \bref{FFst} are given schematically as 

\bea
[\pmatrix{F_a^{~~\bar{b}}\cr \bar{F}_{\bar{a}}^{~~b}},
(F_c^{~~\bar{d}}, \bar{F}_{\bar{c}}^{~~d})]&=&
\pmatrix{ J_{ac}^{\bar{b}\bar{d}}&D_{a\bar{c}}^{d\bar{b}}\cr
D_{\bar{a}c}^{b\bar{d}}&J_{\bar{a}\bar{c}}^{{b}{d}}
}\label{FFst1}\\
J_{ac}^{\bar{b}\bar{d}}&=&{\bf J}_{\langle ac\rangle}\Omega^{\bar{b}\bar{d}}
-\bar{\bf J}^{\langle \bar{b}\bar{d}\rangle}\Omega_{ac}
-\frac{1}{4}{\rm STr}J\Omega_{ac}\Omega^{\bar{b}\bar{d}}\nn\\
 J_{\bar{a}\bar{c}}^{{b}{d}}&=&
 {\bf J}^{\langle {b}{d}\rangle}\Omega_{\bar{a}\bar{c}}
-\bar{\bf J}_{\langle \bar{a}\bar{c}\rangle}\Omega^{bd}
-\frac{1}{4}{\rm STr}J\Omega_{\bar{a}\bar{c}}\Omega^{{b}{d}}\nn\\
D_{a\bar{c}}^{d\bar{b}}&=&
-{\bf D}_a^{~~d}\delta_{\bar{c}}^{\bar{b}}
+\bar{\bf D}_{\bar{c}}^{~~\bar{b}}\delta_{a}^{d}\nn\\
D_{\bar{a}c}^{b\bar{d}}&=&
-{\bf D}_c^{~~b}\delta_{\bar{a}}^{\bar{d}}
+\bar{\bf D}_{\bar{a}}^{~~\bar{d}}\delta_{c}^{b}\nn~~.
\eea 
The right hand side of \bref{FFst1} has half zero modes described symbolically as
\bea
\pmatrix{D&J\cr J& D}\pmatrix{J&D\cr D& J}=
\pmatrix{A_{\parallel}&A_{\perp}\cr A_{\perp}&A_{\parallel}}=0~~%,~~
%\matrix{A_{\perp}=A_++A_-\cr A_{\parallel}=A_+-A_-}
\label{FFAA}~~
\eea
by using the $\tau$ and $\sigma$ reparametrization constraints \bref{AplSUST}
and \bref{AppSUST}.
 As expected the half of the fermionic constraints are first class and another half are 
second class constraints.
The projection matrix into first class is now the one  in \bref{FFAA},
so first class constraints are given by
\bea
\pmatrix{B\cr \bar{B}}\equiv\pmatrix{D&J\cr J& D}\pmatrix{F\cr \bar{F}}
=\pmatrix{\langle{\bf D}\rangle F+F\langle\bar{\bf D}\rangle-
(\langle\bar{\bf J}\rangle \bar{F}+\bar{F}\langle{\bf J}\rangle)^T
\cr
\langle\bar{\bf D}\rangle \bar{F}+\bar{F}\langle{\bf D}\rangle-
(\langle{\bf J}\rangle {F}+{F}\langle\bar{\bf J}\rangle)^T}=0\label{BB}~~.
\eea
They generate following $\kappa$-transformations
\bea
\delta_{\kappa,\bar{\kappa}}Z&=&[Z,{\rm tr}(\bar{B}\kappa-{B}\bar{\kappa})]\label{kappatr}\\
\delta_{\kappa,\bar{\kappa}}X&=&
\Theta(\bar{\kappa}\langle {\bf D}\rangle+\langle \bar{\bf D}\rangle \bar{\kappa}
+{\kappa}^T\langle {\bf J}\rangle+\langle \bar{\bf J}\rangle {\kappa}^T)\nn\\
\delta_{\kappa,\bar{\kappa}}\bar{X}&=&
\bar{\Theta}({\kappa}\langle \bar{\bf D}\rangle+\langle {\bf D}\rangle {\kappa}
+\bar{\kappa}^T\langle \bar{\bf J}\rangle+\langle {\bf J}\rangle \bar{\kappa}^T
)\nn\\
\delta_{\kappa,\bar{\kappa}}\Theta&=&
X({\kappa}\langle \bar{\bf D}\rangle+\langle {\bf D}\rangle {\kappa}
+\bar{\kappa}^T\langle \bar{\bf J}\rangle+\langle {\bf J}\rangle \bar{\kappa}^T
)\nn\\
\delta_{\kappa,\bar{\kappa}}\bar{\Theta}&=&
\bar{X}(\bar{\kappa}\langle {\bf D}\rangle+\langle \bar{\bf D}\rangle \bar{\kappa}
+{\kappa}^T\langle {\bf J}\rangle+\langle \bar{\bf J}\rangle {\kappa}^T
)\nn~~.
\eea
Both $\kappa$ and $\bar{\kappa}$ parameters have following zeromodes,
and this reducibility continues infinitely
\bea
\delta\kappa&=&\langle {\bf D}\rangle {\kappa}_1-{\kappa}_1\langle \bar{\bf D}\rangle
-\langle {\bf J}\rangle \bar{\kappa}_1^{~T}+\bar{\kappa}_1^{~T}\langle \bar{\bf J}\rangle
~,~\delta\kappa_1=\langle {\bf D}\rangle {\kappa}_2~+{\kappa}_2\langle \bar{\bf D}\rangle
+\langle {\bf J}\rangle \bar{\kappa}_2^{~T}+\bar{\kappa}_2^{~T}\langle \bar{\bf J}\rangle,
\cdots\nn\\
\delta\bar{\kappa}&=&\langle \bar{\bf D}\rangle \bar{\kappa}_1-\bar{\kappa}_1\langle {\bf D}\rangle
-\langle \bar{\bf J}\rangle {\kappa}_1^{~T}+{\kappa}_1^{~T}\langle {\bf J}\rangle
~,~\delta\bar{\kappa}_1=\langle \bar{\bf D}\rangle \bar{\kappa}_2~+\bar{\kappa}_2\langle {\bf D}\rangle
+\langle \bar{\bf J}\rangle {\kappa}_2^{~T}+{\kappa}_2^{~T}\langle {\bf J}\rangle
,
\cdots\nn~\\
\label{k0k1k2}
\eea
where $A_{\perp}=0$ and $A_{\parallel}=0$ are used.
Again it leads to that only each half components of $\Theta$ and $\bar{\Theta}$ can be 
fixed by using this $\kappa$-symmetries.

Now we calculate equations of motion using with the Hamiltonian \bref{HamilSUST} 
for an AdS superstring in the conformal gauge $g_{00}+g_{11}=0=g_{01}$,  
\bea
\partial_\tau Z_M^{~~A}=Z_M^{~~B}\Gamma_B^{~~A}
~~,~~\Gamma=
\pmatrix{\langle{\bf D}\rangle&-E^{-1/2}\bar{J}^{T}\cr 
-E^{1/2}J^{T}&\langle\bar{\bf D}\rangle }
\approx
\pmatrix{\langle{\bf D}\rangle&2D\cr 
2\bar{D}&\langle\bar{\bf D}\rangle }
\label{eqZ}~~.
\eea
The momenta satisfies
\bea
&&\partial_\tau \langle{\bf D}\rangle=\nabla_\sigma\langle{\bf J}\rangle~~,~~
\partial_\tau\langle{\bf J}\rangle=\nabla_\sigma\langle{\bf D}\rangle\label{DJeq}\\
&&\partial_\tau \langle\bar{\bf D}\rangle=\nabla_\sigma\langle\bar{\bf J}\rangle~~,~~
\partial_\tau\langle\bar{\bf J}\rangle=\nabla_\sigma\langle\bar{\bf D}\rangle\nn~~,
%\left(\partial_\tau+\pmatrix{0&\nabla_\sigma\cr \nabla_\sigma&0}\right)
%\left(\partial_\tau-\pmatrix{0&\nabla_\sigma\cr \nabla_\sigma&0}\right)
%\pmatrix{\langle{\bf D}\rangle&\langle\bar{\bf D}\rangle\cr 
%\langle{\bf J}\rangle&\langle\bar{\bf J}\rangle}=0~~,
\eea
which is the stringy extension of momentum conservation of the superparticle,
$\dot{\langle{\bf D}\rangle}=0$.
The field equation is given by
\bea
&&
\left\{\right.
-\partial_\tau^2+\partial_\sigma^2+\frac{1}{2}\hat{\Gamma}_+\hat{\Gamma}_-
+\frac{1}{2}\hat{\Gamma}_-\hat{\Gamma}_+\nn\\
&&~
~~
+\left[\hat{\Gamma}_+|_{\rm bose},\hat{\Gamma}_-|_{\rm fermi}\right]
+\left[\hat{\Gamma}_-|_{\rm bose},\hat{\Gamma}_+|_{\rm fermi}\right]
\left.\right\}_B^{~A}
Z_M^{~B}=0~,\nn\\&&~~~\label{eqDJ}\\
&&~~\hat{\Gamma}_\pm
=\pmatrix{\langle\hat{\bf D}_\pm\rangle& 2\hat{\hat{D}}_\pm\cr 
2\hat{\hat{\bar{D}}}_\pm&\langle\hat{\hat{\bar{\bf D}}}_\pm\rangle}\nn
\\
&&
~~\langle\hat{\bf D}\rangle_\pm=\langle{\bf D}\rangle\pm\langle{\bf J}\rangle~~,~~
\langle\hat{\bar{\bf D}}\rangle_\pm=\langle\bar{\bf D}\rangle\pm\langle\bar{\bf J}\rangle~~,~~
\nn\\
&&
~~\hat{\hat{D}}_\pm=D\pm\frac{1}{2}J~~,~~
\hat{\hat{\bar{D}}}_\pm=\bar{D}\pm\frac{1}{2}\bar{J}~~,
\nn
\eea
in a trivial $Sp(4)\times GL(1)$ gauge.
$\hat{\Gamma}_\pm$ are now induced connection of AdS background on
the right/left sectors of the worldvolume.

We have shown that the AdS superstring has the $\tau,\sigma$ reparametrization invariance and
the fermionic constraints that are mixture of first class and second class
as same as the flat superstring.
The field equation for the AdS superstring is given by \bref{eqDJ}
which includes the contribution of the AdS background through the induced connections $\hat{\Gamma}$'s.

It is expected that the solutions are written as sum of right and left moving modes.
Indeed the matrix in the right hand side of \bref{FFst1} can be diagonalized,
where eigenvalues are 
\bea
\hat{\bf D}_{\pm}&=&{\bf D}\pm{\bf J}\label{pmD}\\
\hat{\bar{\bf D}}_{\pm}&=&\bar{\bf D}\pm\bar{\bf J}\nn
\eea
and eigenstates are 
\bea
\hat{D}_{\pm ~a}^{~~\bar{b}}&=&F_a^{~~\bar{b}}\pm\bar{F}_{~~a}^{\bar{b}}
=E^{1/4}\hat{\hat{D}}%(D\pm \frac{k}{2}J)
{}_a^{~~\bar{b}}\pm E^{-1/4}\hat{\hat{\bar{D}}}%(\bar{D}\pm \frac{k}{2}\bar{J})
{}_{~~a}^{\bar{b}}
~~\label{Dpm}.
\eea
%The bosonic constraint generating the $\tau,\sigma$ reparametrization invariance 
%\bref{AplSUST} and \bref{AppSUST}
%are rewritten as
%\bea
%A_{\pm}&=&\frac{1}{4}{\rm tr}[\langle \hat{\bf D}_\pm\rangle ^2 -\langle \hat{\bar{\bf D}}_\pm\rangle ^2
%%]=0~~.
%\eea
The coset part of supercovariant derivative can be separated into right/left moving sectors
up to constraints
\bea
\left[\langle \hat{\bf D}_\pm\rangle_{ab}(\sigma),\langle \hat{\bf D}_\mp\rangle_{cd}(\sigma')\right]&=&
2\Omega_{\langle c\mid \langle b}({\bf D})_{a\rangle\mid d\rangle}\approx 0\nn\\
\left[\langle \hat{\bf D}_\pm\rangle_{ab}(\sigma),\hat{D}_{\mp, c}^{~~\bar{d}}(\sigma')\right]&=&
\Omega_{c\langle b}\hat{D}_{\pm,a\rangle}^{~~\bar{d}}\delta(\sigma-\sigma')\approx 0\nn\\
\left[\hat{D}_{\pm,a}^{~~\bar{b}}(\sigma),\hat{D}_{\mp,c}^{~~\bar{d}}(\sigma')\right]&=&
2\left(\Omega_{ac}(\bar{\bf D})^{\bar{b}\bar{d}}+\Omega^{\bar{b}\bar{d}}({\bf D})_{ac}
\right)\approx
0\label{rlrl}
\eea
and similar algebras for the bared (S$^5$) sector.
This right/left separation corresponds to the choice 
$J_{\rm bose}\to RJ_{\rm bose}$ and $J_{\rm fermi}\to (R/2)J_{\rm fermi}$ in the IW contraction \bref{DpmJe}.
Despite of the general feature of the algebra \bref{DpmJ}, 
this almost chiral representation is possible because of local constraints such as 
bosonic Sp(4) constraints
${\bf D}_{(ab)}=\bar{\bf D}_{(ab)}=0$.
On the other hand this incomplete separation causes the following complicated algebra.
The poisson bracket of supercovariant derivatives
for a $\tau+\sigma$ sector are given as follows, where $_+$ indices are omitted:
\bea
\left[\langle \hat{\bf D}\rangle_{ab}(\sigma),\langle \hat{\bf D}\rangle_{cd}(\sigma')\right]
&=&2\Omega_{\langle c|\langle b}\Omega_{a\rangle|d\rangle}\delta'(\sigma-\sigma')
+4\Omega_{\langle c|\langle b}({\bf J})_{a\rangle|d\rangle}
\delta(\sigma-\sigma')\nn\\
\left[\langle \hat{\bar{\bf D}}\rangle_{\bar{a}\bar{b}}(\sigma),
\langle \hat{\bar{\bf D}}\rangle_{\bar{c}\bar{d}}(\sigma')\right]
&=&-2\Omega_{\langle \bar{c}|\langle \bar{b}}\Omega_{\bar{a}\rangle|\bar{d}\rangle}\delta'(\sigma-\sigma')
+4\Omega_{\langle \bar{c}|\langle \bar{b}}(\bar{\bf J})_{\bar{a}\rangle|\bar{d}\rangle}
\delta(\sigma-\sigma')\nn\\
\left[\hat{D}_a^{~~\bar{b}}(\sigma),\hat{D}_c^{~~\bar{d}}(\sigma')\right]&=&
2(\Omega^{\bar{b}\bar{d}}\langle\hat{\bf D}\rangle_{ac}
-\Omega_{ac}\langle \hat{\bar{\bf D}}\rangle^{\bar{b}\bar{d}}
)\delta(\sigma-\sigma')\nn\\
\left[\langle \hat{\bf D}\rangle_{ab}(\sigma),\hat{D}_c^{~~\bar{d}}(\sigma')\right]&=&
\Omega_{c\langle b}\tilde{D}_{a\rangle}^{~~\bar{d}}\delta(\sigma-\sigma')\nn\\
\left[\langle \hat{\bar{\bf D}}\rangle_{\bar{a}\bar{b}}(\sigma),\hat{D}_c^{~~\bar{d}}(\sigma')\right]&=&
\delta^{\bar{d}}_{\langle \bar{a}}\tilde{D}_{c|\bar{b}\rangle}\delta(\sigma-\sigma')\label{covderst}\\
\left[\tilde{D}_a^{~~\bar{b}}(\sigma),\tilde{D}_c^{~~\bar{d}}(\sigma')\right]&=&
2(-\Omega^{\bar{b}\bar{d}}\langle \hat{{\bf D}}+2{\bf J}\rangle^{ac}
+\Omega_{ac}\langle \hat{\bar{\bf D}}+2\bar{\bf J}\rangle^{\bar{b}\bar{d}})
\delta(\sigma-\sigma')\nn\\
\left[\hat{D}_a^{~~\bar{b}}(\sigma),\tilde{D}_c^{~~\bar{d}}(\sigma')\right]&=&
-4\Omega_{ac}\Omega^{\bar{b}\bar{d}}\delta'(\sigma-\sigma')
-4\{\Omega_{ac}(\bar{\bf J})^{\bar{b}\bar{d}}+
\Omega^{\bar{b}\bar{d}}({\bf J})_{ac})\}
\delta(\sigma-\sigma')
\nn\\
\left[\langle \hat{\bf D}\rangle_{ab}(\sigma),\tilde{D}_c^{~~\bar{d}}(\sigma')\right]&=&
\Omega_{c\langle b}(\hat{D}-2\hat{\omega})_{a\rangle }^{~~\bar{d}}\delta(\sigma-\sigma')\nn\\
\left[\langle \hat{\bar{\bf D}}\rangle_{\bar{a}\bar{b}}(\sigma),\tilde{D}_c^{~~\bar{d}}(\sigma')\right]&=&
\delta^{\bar{d}}_{\bar{\langle \bar{a}}}(\hat{D}-2\hat{\omega})_{c|\bar{b}\rangle}\delta(\sigma-\sigma')\nn
\eea
where
\bea
\tilde{D}_{\pm~a}^{~~\bar{b}}&=&\hat{D}_{\mp~a}^{~~\bar{b}}+2\hat{\omega}_{\pm~a}^{~~\bar{b}}\nn\\
\hat{\omega}_{\pm~a}^{~~\bar{b}}&=&\pm E^{1/4}J_a^{~~\bar{b}}
-E^{-1/4}\bar{J}^{\bar{b}}_{~~{a}}\nn~~.
\eea

Then next we calculate closed first class constraint set
to obtain its local symmetries analogous to the AdS superparticle case.
The first class constraint set, which can be separated into the left/right chiral sectors up to 
the constraints, 
is given by 
\bea
{\cal A}&=&\frac{1}{4}{\rm tr}\left( \langle \hat{\bf D}\rangle ^2 -\langle \hat{\bar{\bf D}}\rangle ^2
+\tilde{D}\hat{D}
\right)\nn\\
{\cal B}_a^{~~\bar{b}}&=&
\langle \hat{\bf D}\rangle_a^{~~{\cdot}} \hat{D}_{\cdot}^{~~\bar{b}}  
+\hat{D}_a^{~~\bar{\cdot}}\langle \hat{\bar{\bf D}}\rangle_{\bar{\cdot}}^{~~\bar{b}}\\
{\cal C}_{a~c}^{~\bar{b}~\bar{d}}&=&\hat{D}_{a}^{~~\bar{b}}\hat{D}_c^{~~\bar{d}}\nn\\
{\cal D}_{\langle ab\rangle}&=& \hat{D}_{\langle a}^{~~\bar{\cdot}}\nabla\hat{{D}}_{{b}\rangle \bar{\cdot}}~~~
,~~~\bar{\cal D}_{\langle \bar{a}\bar{b}\rangle}=\hat{{D}}_{~\langle \bar{a}|}^{{\cdot}}
\nabla\hat{{D}}_{\cdot|\bar{b}\rangle}
~~,\nn
\eea
where indices $\cdot,\bar{\cdot}$ are used as contracted indices.
The number of ${\cal A}$, ${\cal B}$, ${\cal C}$, ${\cal D}$ and $\bar{\cal D}$ are
$1,16,16\times 15/2=120,~5$ and $5$ which coincides with the flat superstring case \cite{WS}.
They satisfy the following algebras up to \bref{sp4gl1}
:
\bea
\left[{\cal A}(\sigma),{\cal A}(\sigma')\right]&=&
2{\cal A}(\sigma)\delta'(\sigma-\sigma')+{\cal A}'\delta (\sigma-\sigma')\nn\\
\left[{\cal A}(\sigma),{\cal B}_c^{~~\bar{d}}(\sigma')\right]&=&
-2{\cal B}_c^{~~\bar{d}}\delta'(\sigma-\sigma')+
\{- \nabla {\cal B}_c^{~~\bar{d}}
+\frac{1}{2}{\cal C}_c^{~~\bar{\cdot}} \langle^{\cdot}_{~~\bar{\cdot}} \hat{\omega}_{-~\cdot}^{~~\bar{d}} \rangle
-\frac{1}{2} \langle \hat{\omega}_{-~c}^{~~\bar{\cdot}} {\cal C}_{~~\bar{\cdot}}^{\cdot} \rangle
^{~~\bar{d}}_{\cdot} \} \delta(\sigma-\sigma')\nn\\
\nabla {\cal B}_c^{~~\bar{d}}&=&
\left({\cal B}'
-2({\bf J}){\cal B}+2{\cal B}(\bar{\bf J})\right)_c^{~~\bar{d}}\nn\\
\left[{\cal A}(\sigma),{\cal C}_{a~c}^{~\bar{d}~\bar{d}}(\sigma')\right]&=&
2{\cal C}_{a~c}^{~\bar{b}~\bar{d}}(\sigma)\delta'(\sigma-\sigma')
+{\cal C}_{a~c}^{'~\bar{b}~\bar{d}}(\sigma)\delta(\sigma-\sigma')\nn\\
\left[{\cal A}(\sigma),{\cal D}_{ab}(\sigma')\right]&=&
-3{\cal D}_{ab}(\sigma)\delta'(\sigma-\sigma')
-2{\cal D}_{ab}'(\sigma)\delta(\sigma-\sigma')\nn\\
\left[{\cal A}(\sigma),\bar{\cal D}_{\bar{a}\bar{b}}(\sigma')\right]&=&
-3\bar{\cal D}_{\bar{a}\bar{b}}(\sigma)\delta'(\sigma-\sigma')
-2\bar{\cal D}_{\bar{a}\bar{b}}'(\sigma)\delta(\sigma-\sigma')\nn\\
%%%%%%%
\left[{\cal B}_a^{~~\bar{b}}(\sigma),{\cal B}_c^{~~\bar{d}}(\sigma')\right]&=&
-\left(
{\cal C}_{\langle a~c\rangle\bar{\cdot}}^{~\bar{\cdot}}
+{\cal C}^{\cdot\langle \bar{b}~\bar{d}\rangle}_{~~~{\cdot}}
+2{\cal C}_{a~c}^{~\bar{d}~\bar{b}}
\right)\delta'(\sigma-\sigma')\nn\\
&&
-2[(\langle \hat{\bf D}\rangle_{ac}\Omega^{\bar{b}\bar{d}}
+\langle \hat{\bar{\bf D}}\rangle^{\bar{b}\bar{d}}\Omega_{ac}){\cal A}\nn\\
&&+\frac{1}{2}({\cal B}_a^{~~\bar{b}}\tilde{D}_c^{~~\bar{d}}
+ {}^a_{\bar{b}}\leftrightarrow {}^c_{\bar{d}})
+({\cal B}\tilde{D}^T)_{\langle ac\rangle}\Omega^{\bar{b}\bar{d}}
-(\tilde{D}^T{\cal B})^{\langle {\bar{b}\bar{d}}
\rangle}\Omega_{ac}
\nn\\
&&-\nabla {\cal C}_{a~c~}^{~\bar{d}~\bar{b}}
+{\cal D}_{ac}\Omega^{\bar{b}\bar{d}}
-\bar{\cal D}^{\bar{b}\bar{d}}\Omega_{ac}
]\delta(\sigma-\sigma')
\nn\\
\nabla {\cal C}_{a~c}^{~\bar{d}~\bar{b}}&=&{\cal C}_{a~c}^{'~\bar{d}~\bar{b}}
	+{\cal C}_{\cdot~*}^{~\bar{\cdot}~\bar{*}}
	\left[\right.\left( 
	({\bf J})_a^{~\cdot}\delta_{\bar{\cdot}}^{\bar{d}}
	+(\bar{\bf J})_{\bar{\cdot}}^{~\bar{d}}\delta_{a}^{\cdot})\delta_c^{*}\delta_{\bar{*}}^{\bar{b}}
	+_a^{\bar{d}}\leftrightarrow_c^{\bar{b}}
	\right)\nn\\
&&	
	-\Omega_{ac}({\bf J})^{\cdot *}\delta_{\bar{\cdot}}^{[\bar{b}}\delta_{\bar{*}}^{\bar{d}]}
	+\Omega^{\bar{b}\bar{d}}(\bar{\bf J})_{\bar{\cdot} \bar{*}}
	\delta_{[a}^{\cdot}\delta_{c]}^{*}
	+({\bf J})_{ac}\delta_{\bar{\cdot}}^{(\bar{b}}\delta_{\bar{*}}^{\bar{d})}\Omega^{*\cdot}
	-(\bar{\bf J})^{\bar{b}\bar{d}}\delta_{(a}^{\cdot}\delta_{c)}^{*}\Omega_{\bar{*}\bar{\cdot}}
	\left.\right]\nn\\
%%%	
\left[{\cal B}_a^{~~\bar{b}}(\sigma),{\cal C}_{c~d}^{~\bar{c}~\bar{d}}(\sigma')\right]&=&
\left[\right.-2\Omega_{ac}\Omega^{\bar{b}\bar{c}}\hat{D}_d^{~\bar{d}}{\cal A}
+\frac{1}{2}\{
(\delta_c^{\cdot}\delta_{\bar{\cdot}}^{\bar{b}}\delta_d^{*}\delta_{\bar{*}}^{\bar{d}}\tilde{D}_a^{~\bar{c}}-
_a^{\bar{c}}\leftrightarrow_c^{\bar{b}})
-(\delta_d^{\cdot}\delta_{\bar{\cdot}}^{\bar{b}}\delta_c^{*}\delta_{\bar{*}}^{\bar{c}}\tilde{D}_a^{~\bar{d}}-
_a^{\bar{d}}\leftrightarrow_d^{\bar{b}})\nn\\&&
-(\delta_{\bar{\cdot}}^{\bar{b}}\delta_d^{*}\delta_{\bar{*}}^{\bar{d}}\tilde{D}^{\cdot \bar{c}}\Omega_{ca}
-_d^{\bar{d}}\leftrightarrow_c^{\bar{c}})
+(\delta_a^{\cdot}\delta_d^{*}\delta_{\bar{*}}^{\bar{d}}\tilde{D}_{c\bar{\cdot}}\Omega^{\bar{c}\bar{b}}
-_d^{\bar{d}}\leftrightarrow_c^{\bar{c}})\nn\\&&
-(\delta_d^{*}\delta_{\bar{*}}^{\bar{d}}\tilde{D}^{\cdot}_{\bar{\cdot}}
\Omega^{\bar{c}\bar{b}}\Omega_{ca}
-_d^{\bar{d}}\leftrightarrow_c^{\bar{c}})
\}{\cal C}_{\cdot~*}^{~\bar{\cdot}~\bar{*}}\left.\right](\sigma)\delta(\sigma-\sigma')
\nn\\
%%%%%%%%%%%%%%%%%%%%%%%%
\left[{\cal B}_a^{~~\bar{b}}(\sigma),{\cal D}_{cd}(\sigma')\right]&=&
\left[\right.
2\hat{D}_c^{~\bar{b}}\Omega_{ad}{\cal A}
+\tilde{D}_d^{~\langle\bar{\cdot}}{\cal C}_{c~a\bar{\bar{\cdot}}}^{~\bar{b}}
-\tilde{D}_c^{~\langle\bar{\cdot}}
{\cal C}_{~\bar{\bar{\cdot}}\langle \cdot}^{\cdot~~~\bar{b}}\Omega_{a\mid d\rangle}
\left.\right]
\delta'(\sigma-\sigma')\nn\\
&&+\left[\right.
4\nabla\hat{D}_c^{~\bar{b}}\Omega_{ad}{\cal A}
-\tilde{D}^{\cdot\bar{\cdot}}\nabla{\cal C}_{\cdot\bar{\cdot}c}^{~~~~\bar{b}}\Omega_{ad}\nn\\
&&
+\tilde{D}_d^{~\bar{\cdot}}\{
-\nabla {\cal C}_{c~a\bar{\cdot}}^{~\bar{b}}
+(\nabla{\cal C}_{c~a\cdot}^{~\cdot}-{\cal D}_{ca})\delta_{\bar{\cdot}}^{\bar{b}}
+(\frac{1}{2}\nabla{\cal C}_{~~\cdot\bar{\cdot}}^{\cdot\bar{b}}+
\bar{\cal D}^{\bar{b}}_{\bar{\cdot}})\Omega_{ac}
\}
\left.\right]
\delta(\sigma-\sigma')\nn\\
&&\label{1stabcd}
\eea
It is noted that the covariant derivatives $\nabla_\sigma$ are not usual covariant derivatives 
on constraints, but $\nabla_\sigma$ acts on parameters as usual $Sp(4)$ covariant derivatives
if one multiply parameters to these algebra.
Since commutators among ${\cal C}$'s and ${\cal D}$'s always contain
two $\hat{D}$'s, they make closed algebras.

The AdS superstring algebra does not have the form of the AdS superparticle algebra \bref{32},
$(D^2)_A^{~B}|_{\rm Traceless}=0$.
Because of the stringy anomalous term 
$[\langle \hat{\bf D}\rangle,\langle \hat{\bf D}\rangle]\sim \delta'(\sigma-\sigma')$,
the poisson bracket between two ${\cal B}$'s contains
$\delta'(\sigma-\sigma')$
 ${\cal C}$ term which  no more has the form of \bref{32}. 
The number of ${\cal C}$ constraints is $120=\left(\begin{array}{c}10\\3\end{array}\right)$ 
which is the number of
 degrees of freedom of three forms in 10-dimensions. 
 The total number of this constraints set is $1+16+120+10=147$ which is not $8^2-1=63$
 of the AdS superparticle case but is same as the flat superstring case \cite{WS,WSTXT}. 
Interesting result is that this first class constraint set ${\cal A}={\cal B}={\cal C}={\cal D}=0$ 
satisfy the same algebra of the one for the flat case \cite{WS,WSTXT}.
This first class constraint set represents the same 
as the system with the second class constraints $F=\bar{F}=0$ in \bref{FFst}.

\par
%%%%%%%%%%%%%%%%%%%%%%%%%%%%%%%%%%%%%%%%%%%%%%%%%%%%
\section{ Conclusions and discussions}\par

In this paper we have examined the manifest superconformal construction 
of the AdS superparticle and superstring described by
the supermatrix valued coordinates.
By introducing supercovariant derivatives whole symmetries of the AdS superparticle
are simply described by $D^2=0$,
adding to $(Sp(4)\times GL(1))^2$ constraints $({\bf D})={\rm tr}{\bf D}=0$. 
$\kappa$-symmetry and its infinitely reducibility for both the AdS superparticle/string 
are shown as that half of each $\Theta$ and $\bar{\Theta}$ can be gauged away. 
So the complex gauge proposed in \cite{RWS} can not be applied.
Equation of motion of the AdS superparticle is obtained as free null AdS geodesics.

The equation of motion for
 the AdS superstring is obtained as two-dimensional wave equation with induced AdS connection.
The left/right chiral separation from this two dimensional equation 
is not obvious, but we obtained left/right chiral supercovariant derivative algebra
from the action.
Chiral separation of the supercovariant derivatives
and spacetime local symmetry algebra are possible 
up to constraints.
The origin of this weakly chiral separation can be explained as that
the AdS superstring is described by supercovariant derivatives and currents 
which satisfy inhomogeneous $GL(4|4)$ rather than just $GL(4|4)$.
In other words stringy extension is not incorporated with the superconformal symmetry
$GL(4|4)$.
Local symmetry constraint set for the AdS superstring
 obtained by direct computation from the action 
is turned out to be the same with the one for the flat case, ${\cal ABCD}$ constraints.

Quantization of these systems are possible in the first class constraints approach 
for the AdS superparticle and the ``light-cone-like" gauge for both the AdS superparticle and string
as we have shown.
This model is easy to handle as the AdS superstring system,
and will be useful toward the AdS superstring correction to the classical AdS supergravity theory.

%\end{document}
\vskip 6mm
{\bf Acknowledgments}\par
This research is partially supported by the grant-in-aid for Scientific 
Research, \\\noindent
No.12640258 Ministry of Education Japan.
%\appendix
%\eject
%%%%%%%%%%%%%%%%%%%%%%%%%%%%%%%%%%%%%%%%%%%%%%%%%%%%%%%%%%%%%%%%%%


\vspace{0.5cm}
\begin{thebibliography}{99}
\bibliographystyle{unsrt}
%
\setlength{\itemsep}{0.0in}
\bibitem{Mal} 
L.Maldacena, Adv. Theor. Math. Phys. {\bf 2} (1998) 231, hep-th/9711200;\\
S.S.Gubser, I.R.Klebanov  and A.M.Polyakov, Phys.Lett. {\bf B248} (1998) 105, hep-th/9802109;
\\
E.Witten, Adv. Theor. Math. Phys. {\bf 2} (1998) 253, hep-th/9802150.
\bibitem{MeTsy} 
R.R.Metsaev and A.A.Tseytlin, Nucl. Phys. {\bf B533} (1998) 109, hep-th/9805028;\\
Phys. Lett. {\bf B436} (1998) 281, hep-th/9806095. 
\bibitem{Ram} 
R.Kallosh, J.Rahmfeld and A.Rajaraman, J. High Energy Phys. {\bf 9809} (1998) 002, hep-th/9805217;\\
R.Kallosh and J.Rahmfeld, Phys. Lett. {\bf B443} (1998) 143, hep-th/9808038;\\
J.Rahmfeld and A.Rajaraman, Phys. Rev. {\bf D60} (1999) 64014,  hep-th/9809164;\\
I.Pesando, J. High Energy Phys. {\bf 9811} (1998) 002, hep-th/9808020;\\
% J. High Energy Phys. 02 (1999) 007, hep-th/9809145;\\
%J.Park and S-J.Rey, J. High Energy Phys. 01 (1999) 001, hep-th/9812062;\\
%J-G. Zhou, Nucl. Phys. {\bf B559} (1999) 92, hep-th/9906013.
R.Kallosh and A.A.Tseytlin, J. High Energy Phys. {\bf 9810} (1998) 016 hep-th/9808088.
\bibitem{RamQ}
 A.Rajaraman and M.Rozali,  Phys. Lett. {\bf B468} (1999) 58, hep-th/9902046;\\
 S.Forste, D.Ghoshal and S.Theisen, J. High Energy Phys. {\bf 9908} (1999) 0013, hep-th/9903042;\\
N.Drukker, D.J.Gross and A.A.Tseytlin, J. High Energy Phys. {\bf 0021} (2000) 0004, hep-th/0001204. 
\bibitem{MeTlc} R.R.Metsaev,  Class. Quant. Grav.{\bf 18}(2001) 1245, hep-th/0012026;\\
R.R.Metsaev, C.B.Thorn and A.A.Tseytlin, Nucl. Phys. {\bf B 596} (2001) 151, hep-th/000917.
\bibitem{RWS} R.Roiban and W.Siegel, J. High Energy Phys. {\bf 0011} (2000) 024, hep-th/0010104.
\bibitem{WS} W.Siegel, Nucl. Phys. {\bf B263} (1985) 93.
\bibitem{WS88} W.Siegel, Phys. Lett. {\bf B203} (1988) 79.
\bibitem{WSTXT} 
W.Siegel, ``{\it Introduction to string field theory}" (World Scientific, Singapore, 1988).
\bibitem{WSMH} W.Siegel, ``Free field equation for everything", in string , Cosmology, Composite Structures (March 11-18,1987, College Park . Mryland), eds. S.T.Gates Jr. and R.N.Mohapatra (World Scientific, Singapore, 1987).
\bibitem{WSMH2} M.Hatsuda, Int. J. Mod. Phys. {bf A7} (1992) 1187. 
\bibitem{IW}  E.In$\ddot{\rm o}$n$\ddot{\rm u}$ and E.P.Wigner, Proc. Natl. Acad. Sci. U.S.A. {\bf 39} (1953) 510,\\
E.In$\ddot{\rm o}$n$\ddot{\rm u}$: in ``Group Theoretical Concepts  and Methods in Elementary Particle Physics'' ed. F.Gursey (Gordon and Breach, New York, 1964).
\bibitem{HrKm} M.Green and J.H.Schwarz, Nucl. Phys. {\bf B 243} (1984) 285;\\
T.Hori and K.Kamimura, Prog. Theor. Phys. {\bf 73} (1985) 476.
\bibitem{KT} K.Kamimura and M.Tatewaki (Hatsuda),  Phys. Lett. {\bf B205} (1988) 257;\\
 A.R.Mikovi$'{\rm c}$ and W.Siegel, Phys. Rev. {\bf B209} (1988) 47.
\bibitem{ABCD} F.Essler, E.Laenen, W.Siegel and J.P.Yamron, Phys. Lett. {\bf B254} (1991) 411;\\
F.Essler, M.Hatsuda, E.Laenen, W.Siegel and J.P.Yamron, Nucl. Phys. {\bf B364} (1991) 67.
%%\bibitem{AdSQ} M.Hatsuda and K.Kamimura, in preparation.
\par
\end{thebibliography}
\end{document}